\documentclass[a4paper,12pt]{article}
\usepackage{graphicx} 
\usepackage{dcolumn}
\usepackage{graphicx}
\usepackage{amsmath}
\usepackage{amsfonts}
\usepackage{amssymb}
\usepackage{psfrag}
\usepackage{wrapfig}
\usepackage{subfigure}
\usepackage{makeidx}
\usepackage{bm}
\usepackage{epsf}
\usepackage{color}
\usepackage{cases}
\usepackage{multirow}
\usepackage{booktabs} 
\usepackage{makecell} 
\usepackage{float}
\usepackage{comment}

\newcommand{\be}{\begin{equation}}
\newcommand{\ee}{\end{equation}}
\newcommand{\bea}{\setlength\arraycolsep{2pt} \begin{eqnarray}}
\newcommand{\eea}{\end{eqnarray}}
\newcommand{\nn}{\nonumber}

\usepackage{authblk}

\usepackage{color}
\usepackage[normalem]{ulem}
\usepackage[dvipsnames]{xcolor}
\usepackage{cite}
\usepackage{hyperref}

\def\ft#1#2{{\textstyle{\frac{\scriptstyle #1}{\scriptstyle #2} } }}

\def\0{{\sst{(0)}}}
\def\1{{\sst{(1)}}}
\def\2{{\sst{(2)}}}
\def\3{{\sst{(3)}}}
\def\4{{\sst{(4)}}}
\def\5{{\sst{(5)}}}
\def\6{{\sst{(6)}}}
\def\7{{\sst{(7)}}}
\def\8{{\sst{(8)}}}
\def\sst#1{{\scriptscriptstyle #1}}

\title{Phase Structure of Scalarized Black Holes in Einstein-Scalar-Gauss-Bonnet Gravity}

\author[1,4]{Carlos Herdeiro\thanks{\href{mailto:herdeiro@ua.pt}{herdeiro@ua.pt}}}

\author[2]{Hyat Huang\thanks{\href{mailto:hyat@mail.bnu.edu.cn}{hyat@mail.bnu.edu.cn}}}
 
\author[3]{Jutta Kunz\thanks{\href{mailto:jutta.kunz@uni-oldenburg.de}{jutta.kunz@uni-oldenburg.de}}} 

\author[2]{Meng-Yun Lai\thanks{\href{mailto:mengyunlai@jxnu.edu.cn}{mengyunlai@jxnu.edu.cn}}}  

\author[1]{Eugen Radu\thanks{\href{mailto:eugen.radu@ua.pt}{eugen.radu@ua.pt}}}

\author[2]{De-Cheng Zou\thanks{\href{mailto:dczou@jxnu.edu.cn}{dczou@jxnu.edu.cn}}}

\affil[1]{Departamento de Matem\'atica da Universidade de Aveiro and
Centre for Research and Development in Mathematics and Applications (CIDMA),
Campus de Santiago, 3810-183 Aveiro, Portugal}
\affil[2]{College of Physics and Communication Electronics, Jiangxi Normal University, Nanchang 330022, China}
\affil[3]{Institut f\"ur  Physik, Universit\"at Oldenburg, Postfach 2503,
D-26111 Oldenburg, Germany}
\affil[4]{Programa de P\'os-Gradua\c{c}\~{a}o em F\'{\i}sica, UFPA, 66075-110, Bel\'em, Brazil.}

\date{March 2026}

\begin{document}

\maketitle

\begin{abstract}
We revisit scalarized black holes in Einstein–scalar–Gauss–Bonnet gravity and analyze the thermodynamic phase transition between the Schwarzschild solution of general relativity and scalarized black holes. Restricting to spherically symmetric configurations, we investigate several classes of scalar-Gauss-Bonnet coupling functions. For the simplest quadratic coupling that triggers spontaneous scalarization, the scalarized solutions are thermodynamically disfavored and no phase transition occurs. For an exponential coupling, the phase structure depends strongly on the coupling parameter, allowing for the absence of a transition, a continuous second-order transition, or a discontinuous first-order transition. For couplings leading to purely nonlinear scalarization, we find either a first-order transition or no transition. These results reveal a rich phase structure of scalarized black holes controlled by the scalar–Gauss–Bonnet coupling.
\end{abstract}


\section{Introduction}

As realized long ago by Damour and Esposito-Far\`ese \cite{Damour:1993hw}, compact neutron stars may exhibit the phenomenon of matter-induced spontaneous scalarization in scalar–tensor theories (STTs) \cite{Jordan:1949zz,Fierz:1956zz,Jordan:1959eg,Brans:1961sx,Dicke:1961gz}.
For this phenomenon to occur, the coupling function must be chosen such that general relativistic neutron stars develop a tachyonic instability of the scalar field.
Recently, matter-induced spontaneous scalarization of neutron stars has been revisited, and it was shown that the transition between general relativistic and scalarized neutron star solutions is predominantly of first order \cite{Unluturk:2025zie,Muniz:2025egq}.
Subsequently, phase transitions of scalarized boson stars in STTs were investigated, revealing that the order of the transition depends sensitively on the boson self-interaction potential \cite{Huang:2025dgc}.

\medskip

In Einstein-scalar-Gauss-Bonnet (EsGB) gravity, the vacuum Schwarzschild and Kerr black holes (BHs) of general relativity may also undergo spontaneous scalarization \cite{Doneva:2017bvd,Silva:2017uqg,Antoniou:2017acq,Antoniou:2017hxj,Blazquez-Salcedo:2018jnn,Silva:2018qhn,Macedo:2019sem,Cunha:2019dwb,Collodel:2019kkx,Macedo:2020tbm,Blazquez-Salcedo:2020rhf,Blazquez-Salcedo:2020caw,Dima:2020yac,Hod:2020jjy,Herdeiro:2020wei,Berti:2020kgk}.
In this case the scalarization mechanism is curvature-induced, arising from the coupling of the scalar field to the Gauss-Bonnet invariant.
Above a certain curvature threshold the Schwarzschild or Kerr solution develops a tachyonic instability, which is subsequently quenched by nonlinear effects and leads to a new branch of scalarized black hole solutions.
However, scalarization of black holes need not rely on such a tachyonic instability.
If the coupling function contains higher-order terms in the scalar field, scalarized solutions may also arise through purely nonlinear effects, leading to the phenomenon of nonlinear scalarization \cite{Doneva:2021tvn,Blazquez-Salcedo:2022omw,Doneva:2022yqu}.
Moreover, numerical relativity simulations of black hole binaries have shown that dynamical scalarization and descalarization may occur during the inspiral or merger of black holes (see, $e.g.$, \cite{Silva:2020omi,Doneva:2022byd,Elley:2022ept}).

\medskip

These results indicate that the precise form of the scalar-Gauss-Bonnet coupling function plays a crucial role in determining the scalarization mechanism and the properties of the resulting solutions.
The main purpose of this work is to investigate the thermodynamic phase transition between the Schwarzschild black holes of general relativity and scalarized black holes in EsGB gravity, and to clarify how the existence and the order of this transition depend on the coupling function.
To this end, we analyze the thermodynamic properties of scalarized black holes for three representative classes of coupling functions.
Depending on the choice of coupling and its parameters, we find three possible scenarios: the absence of a phase transition, a continuous second-order transition, or a discontinuous first-order transition.
As a first exploratory step, we restrict our analysis to spherically symmetric black holes.

\medskip

The paper is organized as follows.
In Sec.~2 we briefly review the theoretical framework and discuss the radial stability of Schwarzschild
black hole together with the thermodynamic quantities relevant for the phase transition analysis.
In Sec.~3 we present the resulting phase structure for the different coupling functions.
Finally, Sec.~4 summarizes our conclusions and outlines possible directions for future work.

\section{Theoretical Settings}
\label{Theory}
%
\subsection{Einstein-scalar-Gauss-Bonnet theory}

We consider the action of EsGB theory
\be
\label{M1}
S_\text{EsGB} =\int \sqrt{-g}\bigg(R-\ft{1}{2}(\partial \phi)^2+\alpha f(\phi)R^2_{GB}\bigg) ,
\ee
where $R$ is the Ricci scalar, and $\phi$ is the scalar field that is coupled to the Gauss-Bonnet term $R^2_{GB}$
\be
R^2_{GB}=R^2-4R_{\mu\nu}R^{\mu\nu}+R_{\mu\nu\rho\sigma}R^{\mu\nu\rho\sigma},
\ee
via the coupling function $f(\phi)$ with coupling constant $\alpha$.

Denoting the Einstein tensor by $G_{\mu\nu}=R_{\mu\nu}-\ft{1}{2}g_{\mu\nu}R$, the equations of motion (EOMs) are obtained by the variation principle,
\bea
\label{eom}
&&G_{\mu\nu}+\alpha H_{\mu\nu}=T^\phi_{\mu\nu}, \\
&&\nabla^\mu\nabla_\mu \phi+\alpha\frac{df(\phi)}{d\phi} R^2_{GB}=0,
\eea
with 
\bea
H_{\mu\nu}&=&-R(\nabla_\mu \Psi_\nu+\nabla_\nu\Psi_\mu)-4\nabla^\rho\Psi_\rho G_{\mu\nu}+4 R_{\mu\rho}\nabla^\rho\Psi_\nu+4R_{\nu\rho}\nabla^\rho\Psi_\mu\nn\\
&-&4g_{\mu\nu}R^{\rho\sigma}\nabla_\rho\Psi_\sigma+4R^\sigma_{\ \mu\rho\nu}\nabla^{\rho}\Psi_\sigma,\nn\\
T^\phi_{\mu\nu}&=&\ft{1}{2}(\partial_\mu \phi)(\partial_\nu \phi)-\ft{1}{4}g_{\mu\nu}(\partial \phi)^2,
\eea
where we define $\Psi_\mu=\ft{\partial f(\phi)}{\partial\phi}\nabla_\mu\phi$.

One can easily verify that the Schwarzschild  BH is a solution of
the equations (\ref{eom}) with $\phi\equiv 0$
provided that $\frac{\partial f(\phi)}{\partial\phi}\big|_{\phi=0}=0$.

\medskip

In what follows we here consider the following three choices\footnote{
In addition to (\ref{couplings}), (\ref{couplings-nl}), we have considered as well the  
coupling functions
\begin{eqnarray}
f(\phi)=\frac{\phi^2}{8} -\frac{\beta \phi^4}{64} ~~~{\rm and}~~
f(\phi)= \frac{\phi^4}{64}-\beta\frac{\phi^8}{2048} , 
\end{eqnarray}
(again with $\beta>0$)
which can be taken as the 4th order small-$\phi$
expansion of the type (ii) and type (iii) functions, respectively.
Since the results in these cases are 
qualitatively
similar to those found for exponential couplings couplings,
we shall not reported them here.
}
 for the  coupling function $f(\phi)$ 
  (recall that the coupling functions are multiplied by the coupling constant $\alpha$)
:
\begin{eqnarray}
\label{couplings}
&&
 {\rm {\bf spontaneous\:scalarization \:}} - 
\begin{cases}
{\rm type \:(i)\:function}  : \:  f(\phi)=\frac{\phi^2}{8} +\frac{\beta \phi^4}{64}, \quad \beta \ge 0  , 
\\
{\rm type \:(ii)\:function}   : \:f(\phi)=\frac{1}{2\beta}
    \left[1-\exp(-\frac{\beta\phi^2}{4})\right] , \: \beta \ge 0  , 
\end{cases}
\\
\label{couplings-nl}
&&
{\rm  {\bf non-linear\: scalarization }}-
\: {\rm type \:(iii)\:function}  : \:  f(\phi)=\frac{1}{4\beta}\left[1-\exp(-\frac{\beta\phi^4}{16})\right] ,   \:  \beta > 0 .
\end{eqnarray}
 
 These choices correspond to different situations.
 For {\it spontaneous scalarization},
 the contribution of the  coupling function  $f(\phi)$ 
 results in a tachyonic mass term 
 (sourced by the GB term)
in  the equation  for the 
 linearized  scalar perturbation.  
On the other hand, the type (iii) coupling function  
does not yield a mass term in the linearized Klein-Gordon equation and therefore only  
non-linearly scalarized BHs  may exist. 
 We remark that choosing a  coupling function 
which allows for spontaneous scalarization
does not exclude {\it a priori} the existence of 
branches of scalarized solutions disconnected from Schwarzschild BH, which indeed is the case for a type (ii) coupling function, as we shall see. 

\medskip

Throughout this paper, we consider the static and spherically symmetric ansatz 
\be
ds^2=-A(r)dt^2+B(r)^{-1}dr^2+r^2(d\theta^2+\sin^2\theta d\varphi^2)
\label{metric}
\ee
for the spacetime and 
\be
\phi \equiv \phi(r)
\ee
for the scalar field.
The EsGB equations \eqref{eom} result in a set of
ordinary differential equations which are solved numerically subject to the boundary conditions that the solutions are asymptotically flat and regular on the horizon.
For the employed metric ansatz, the horizon is located at $r=r_H>0$,
with $A(r_H)=B(r_H)=0$.
Scalarized solutions exist only if the expansion at the horizon of the scalar field leads to a non-negative radicand denoted by $\Delta$ 
\cite{Doneva:2017bvd}
\begin{equation}
    \Delta=\left (1-\frac{96\alpha^2}{r^4 }\bigg(\frac{df(\phi)}{d\phi}\bigg)^2 \right) \bigg |_{r=r_H} .
    \label{Delta}
\end{equation}

\subsection{Radial (in)stability of Schwarzschild}

We briefly address the radial (in)stability of the Schwarzschild BH
when viewed as a solution of the model (\ref{M1}).
To this end, we consider the scalar field perturbation
\begin{equation}
\delta \phi(t,r,\theta,\varphi)=\frac{u(r)}{r}e^{-i\omega t}Y_{\ell m}(\theta,\varphi),
\end{equation}
in a background with $A(r)=B(r)=1-2M/r$,
and introduce a tortoise coordinate $dr_*=dr/(1-2M/r)$.
In the case of spontaneous scalarization, this leads to the radial equation \cite{Doneva:2017bvd}
\begin{equation}
\label{pert}
\frac{d^2u}{dr_*^2}+\Big[\omega^2-U(r)\Big]u(r)=0,
\end{equation}
where the potential $U(r)$ is given by
\begin{equation} 
\label{pot-c}
U(r)=\Big(1-\frac{2M}{r}\Big)\left(\frac{\ell (\ell+1)}{r^2}+\frac{2M}{r^3}-\frac{12\alpha M^2}{r^6}\right) \: .
\end{equation}
For a type (iii) coupling function, however, there is no contribution from the coupling function to the above potential and thus the solutions for (\ref{pert})
are those of the usual Klein-Gordon equation.

For {\it spontaneous scalarization}, the Schwarzschild BH develops a first tachyonic instability in the radial $(\ell=0)$-mode at the critical threshold value $ \tilde M\equiv  M/\sqrt{\alpha} \simeq 0.587$,
where Eq.~\eqref{pert} has the eigenvalue $\omega=0$.
Such a zero mode occurs in fact for a whole sequence of values $\tilde{M}_n$, corresponding to (spherical) scalarized solutions with $n=0,1, \dots$ nodes in the scalar field profile.
From each of these values $\tilde{M}_n$ an unstable mode of the Schwarzschild BH emerges \cite{Blazquez-Salcedo:2018jnn}.
Thus for masses $\tilde M < \tilde{M}_0 \simeq 0.587$ the Schwarzschild BH is unstable. 
The eigenvalue $\omega^2$ is real and negative, $i.e.$, $\omega^2 =(i \omega_i)^2$. 

For a scalar coupling function with
$\frac{\partial^2 f(\phi)}{\partial\phi^2}\big|_{\phi=0}=0$,
the Schwarzschild BHs do not develop a tachyonic instability and remain linearly stable.
However, this does not exclude the existence of scalarized solutions
which are not continuously connected to it -- the so-called  {\it non-linear scalarization}.
In fact, as we shall see, such solutions may exist as well for 
a type (ii) coupling function, in which case
$\frac{\partial^2 f(\phi)}{\partial\phi^2}\big|_{\phi=0}\neq 0$.

\subsection{Quantities of interest}
The mass $M$ of the BHs and the scalar charge $Q_s$ can be read from the asymptotic behavior of the functions $A(r)$ and $\phi(r)$,
\be
\label{inf}
-g_{tt}|_{r\to\infty}=A(r)|_{r\to\infty}=1-\frac{2M}{r}+\cdots,\qquad
\phi(r)= 
\frac{Q_s}{r}+\cdots .
\ee
One remarks the system is invariant under the change of sign of $\phi$,
which translates into $Q_s\to -Q_s$.

The Hawking temperature $T_H$ of the BHs is defined via their surface gravity $\kappa$.
For the spherically symmetric line element (\ref{metric}) this results in \cite{Torii:1993vm}
\begin{equation}
\label{THS}
   T_H = \frac{A'(r_H)}{4\pi} \sqrt{\frac{B(r_H)}{A(r_H)}} .
\end{equation}
For the entropy $S$ of EsGB BHs, we employ Wald's entropy
\cite{Wald:1993nt,Iyer:1994ys} which  in the spherically symmetric case yields \cite{Doneva:2017bvd} 
\be
S=\pi r_H^2+4\pi \alpha f(\phi_H) .
\ee
The scalarized solutions satisfy the 1st law 
\be
dM=T_H dS
\ee
and the Smarr relation
\cite{Cunha:2019dwb}
(see also \cite{Ballesteros:2025wvs})
\be
M=2 T_H S+M_{(\phi)} \: \:  \: {\rm with  }\:  \:  \:
M_{(\phi)}=\frac{1}{2 }
\int_{r_H}^\infty  dr \frac{f(\phi)}{\dot f(\phi) } 
\frac{d}{dr} \left (\sqrt{AB} r^2 \frac{d\phi}{dr} \right),
\ee
where $\dot f(\phi)=df(\phi)/d\phi$.

Finally, to fix a residual scaling symmetry of the system, we define dimensionless quantities  in units set by the coupling constant 
$\alpha$ (with $F$ the free energy $cf.$ (\ref{F}))
\be
\label{scaling}
\tilde M=\frac{M}{\sqrt{\alpha}}, ~~\tilde Q_s=\frac{Q_s}{\sqrt{\alpha}},  ~~\tilde T_H=T_H \sqrt{\alpha}, ~~\tilde S=\frac{S}{ {\alpha}} ~~{\rm and}~\tilde F=\frac{F}{\sqrt{\alpha}}~.
\ee

\subsection{Phase transitions}
The scalarization of BHs can be interpreted as a phase transition between two equilibrium configurations: the unscalarized Schwarzschild BH of general relativity and a scalarized EsGB BH with a nontrivial scalar field.
As we shall see, depending on the choice of the coupling function and the values of its parameters, this transition may be continuous (second order), discontinuous (first order), or may not occur at all.

Following the phenomenological approach of Unl\"ut\"urk \textit{et al.}~\cite{Unluturk:2025zie} and Huang \textit{et al.}~\cite{Huang:2025dgc} for neutron stars and boson stars, we adopt a Landau-type framework to classify the phase transitions.
For horizonless compact objects, the Arnowitt-Deser-Misner mass can be expanded in terms of an order parameter such as the scalar charge or the central value of the scalar field.
Stable (metastable) configurations then correspond to global (local) minima of the binding energy.

For BHs in the canonical ensemble, however, the relevant thermodynamic potential is the free energy\footnote{The free energy can be obtained from the on-shell Euclidean action $I_E$ of a BH, identifying $F=T_H I_E$ as a saddle-point approximation.
In this approach the action (\ref{M1}) is supplemented by the Gibbons-Hawking-York boundary term \cite{Gibbons:1976ue,York:1972sj} together with a scalar-Gauss-Bonnet contribution \cite{Julie:2020vov}.}
\be
\label{F}
F = M - T_H S ,
\ee
where $M$, $T_H$ and $S$ are computed from (\ref{inf}) and (\ref{THS}).
For the Schwarzschild solution (denoted by the index ``S'') one has
\be
\label{FS}
M_S=\frac{1}{8\pi T_H}, \qquad
S_S=\frac{1}{16\pi T_H^2}, \qquad
F_S=\frac{1}{16\pi T_H}.
\ee
The corresponding quantities for scalarized solutions are obtained from the numerical data.
For a given temperature, the thermodynamically preferred configuration is the one with the lowest free energy.

In this framework the Hawking temperature $T_H$ plays the role of a {\it control parameter}.
A natural {\it order parameter} for the scalarized phase is the scalar charge $Q_s$, extracted from the $1/r$ falloff of the scalar field at spatial infinity (cf.\ Eq.~(\ref{inf})). This quantity vanishes identically for the Schwarzschild solution and acquires a nonzero value for scalarized configurations, thus capturing the breaking of the 
$\phi\rightarrow  -\phi$ symmetry. 
Equivalently, in spherical symmetry one may also use the horizon value $\phi_H$ of the scalar field.

Within Landau theory, the free-energy difference between scalarized and Schwarzschild black holes can be expanded in powers of the order parameter,
\be
\Delta F(T_H;Q_s) \equiv F_{\text{EsGB}} - F_{\text{S}}
= a(T_H) Q_s^2 + \frac{1}{2} b Q_s^4 + \frac{1}{3} c Q_s^6 + \cdots ,
\label{le}
\ee
where only even powers appear due to the $\phi \rightarrow -\phi$ symmetry of the system.
If $b>0$, the transition is second order: when $a(T_H)$ changes sign, the minimum of $\Delta F$ shifts continuously from $Q_s=0$ to $Q_s\neq0$, and the scalar charge grows smoothly from zero.
In this case the entropy (the first derivative of $F$ with respect to temperature) remains continuous, whereas the second derivative of $F$ is discontinuous or divergent.

If $b<0$ and $c>0$, the transition becomes first order.
The function $\Delta F$ develops two local minima separated by a barrier, allowing the scalarized and unscalarized configurations to coexist as locally stable states over a range of parameters.
At the transition point the global minimum changes discontinuously, leading to a jump in both the scalar charge and the entropy.
For a phase transition to occur toward the thermodynamically preferred state, the corresponding solution should also be linearly stable and hyperbolic.

\medskip

This framework provides a convenient way to interpret the phase structure revealed by our numerical solutions for different coupling functions $f(\phi)$ and coupling parameters $\beta$, including the appearance of metastable branches and changes in the order of the transition.
It also offers a unified thermodynamic perspective on scalarization in different compact objects: while neutron stars and boson stars are compared through their binding energy, BHs are naturally compared through their free energy.

Finally, we note that most of the scalarized BHs considered in this work are not {\it locally} stable in the canonical ensemble, since they possess a negative specific heat, a feature shared with the Schwarzschild solution.
Canonical equilibrium could in principle be restored by placing the system in a finite cavity and fixing the boundary temperature.
Alternatively, one may consider the microcanonical ensemble, where configurations with the same mass are compared through their entropy (with $S=4\pi M^2$ for the Schwarzschild solution). In that case the thermodynamically preferred configuration is the one with the larger entropy.  While this viewpoint can lead to different thermodynamic preferences, in this work we shall primarily rely on the canonical ensemble.

 We emphasize that thermodynamic preference, as determined by the free energy, does not necessarily imply dynamical (linear) stability. In particular, a configuration with lower free energy may still possess unstable perturbative modes. Therefore, in the discussion below, both thermodynamic considerations and the available results on linear stability must be taken into account when assessing the physical relevance of the solutions.

\section{Results}

For each choice of the  coupling function  $f(\phi)$ in (\ref{couplings}), we have
constructed the scalarized BHs by using standard numerical methods,
our results being in full agreement with those reported in the literature.
Therefore  we shall not insist here on such aspects, focusing 
instead on the thermodynamical properties of the solutions,
with the aim of establishing the absence or presence of a phase transition together with its order.

\subsection{Spontaneous scalarization } 

A distinguishing feature of the 
  spontaneous  scalarization
  is the presence  of
 a critical value $ T_H^{(c)}$ of the control parameter $T_H$
 (with  $\tilde T_H^{(c)} \simeq 0.0678$),
 at which the linearized scalar equation on the Schwarzschild background admits a \emph{static normalizable zero mode}. 
As discussed above,
this zero mode marks the threshold of a tachyonic instability 
 \cite{Doneva:2017bvd,Silva:2017uqg}.
There a new family of BH solutions with nontrivial scalar profile branches off from the Schwarzschild solution. 
  At the bifurcation point the order parameter vanishes smoothly,
 $Q_s\to 0$. 
 The set of scalarized BHs can be viewed as a non-linear continuation
 of the zero mode.
The bifurcation point with the Schwarzschild solution is universal, since it requires
only the presence of a $\phi^2$-term in the small-$\phi$ expansion
of the coupling function, the higher order terms being irrelevant.
However, the details of the scalarized solutions, in particular
their thermodynamical features, depend on the 
 choice of the coupling function together with the parameters inside.
 In terms of the Landau expansion~\eqref{le}, the bifurcation point corresponds to the temperature at which the coefficient 
$a(T_H)$ changes sign, leading to the onset of a nontrivial order parameter $Q_s$. 

\subsubsection{A type (i) function: 
$f(\phi)=\frac{1}{8}\phi^2+\frac{\beta \phi^4}{64}$}

 \begin{figure}[ht!]
 	\begin{center}
    \includegraphics[height=.34\textwidth, angle =0 ]{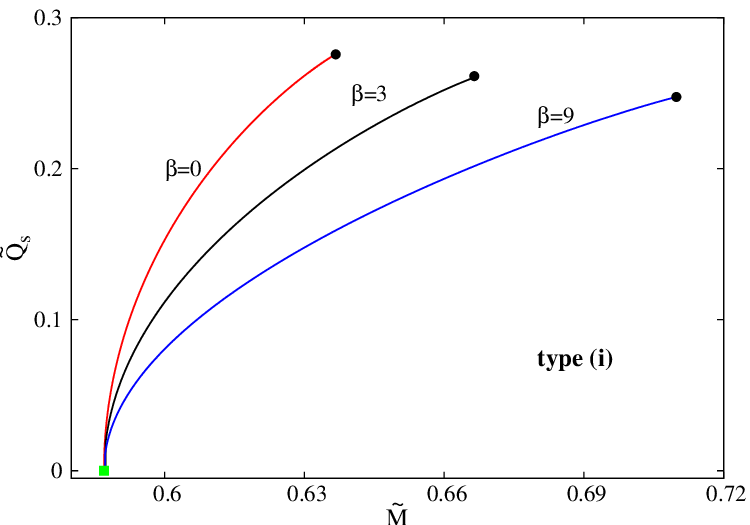}
    \includegraphics[height=.34\textwidth, angle =0 ]{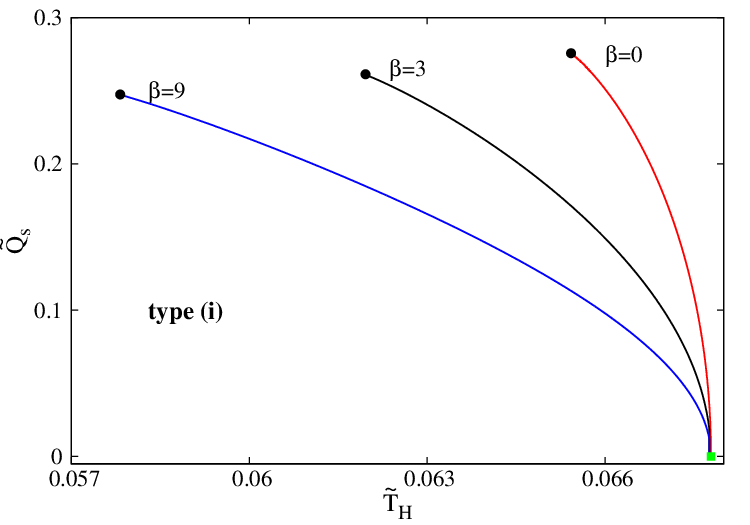} 
 	\end{center}
 	\caption{  
 		{\small 
 		The scalar charge is shown as a function of the BH mass and
        as a function of the Hawking temperature.
  Here and in Figures \ref{type1-2}, \ref{type1-3}
  a type (i) scalar field coupling function,
  $f(\phi)=\frac{1}{8}\phi^2+\frac{\beta \phi^4}{64}$ is considered.
       In all plots, the quantities are given in units set by the coupling
 		}
 	}
 	\label{type1-1}
 \end{figure}

 \begin{figure}[ht!]
 	\begin{center}
    \includegraphics[height=.34\textwidth, angle =0 ]{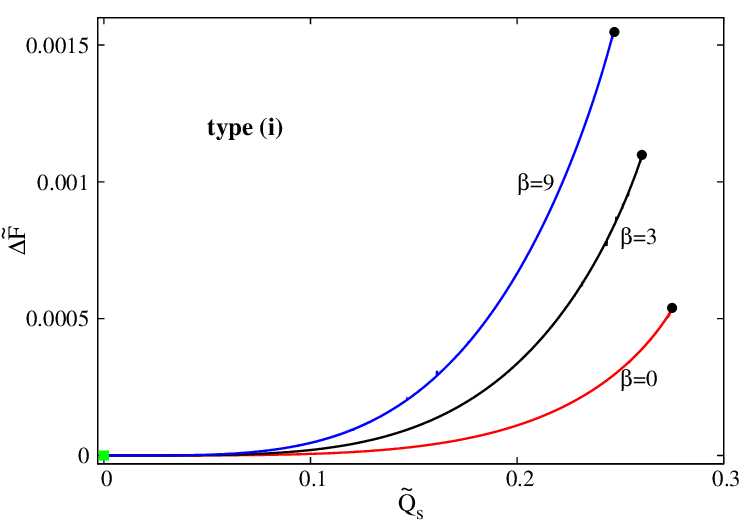}
    \includegraphics[height=.34\textwidth, angle =0 ]{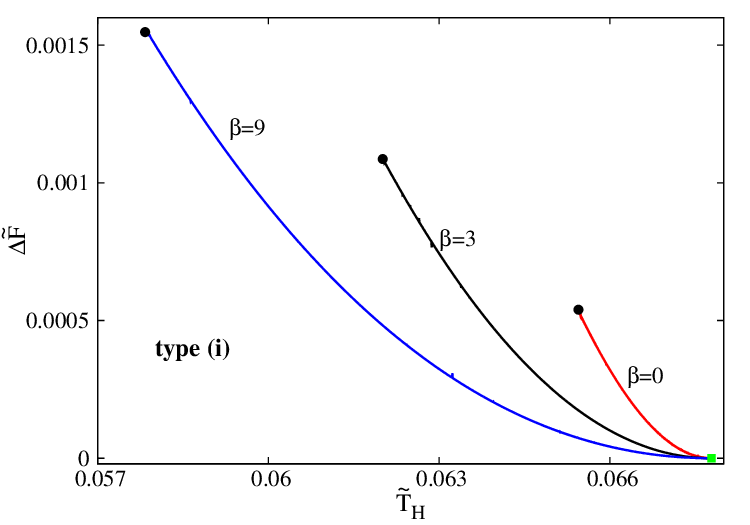} 
 	\end{center}
 	\caption{  
 		{\small 
 		The difference between the free energy of scalarized and Schwarzschild BHs 
       is shown as a function of mass and  Hawking temperature.
 		}
 	}
 	\label{type1-2}
 \end{figure}

 \begin{figure}[ht!]
 	\begin{center}
    \includegraphics[height=.34\textwidth, angle =0 ]{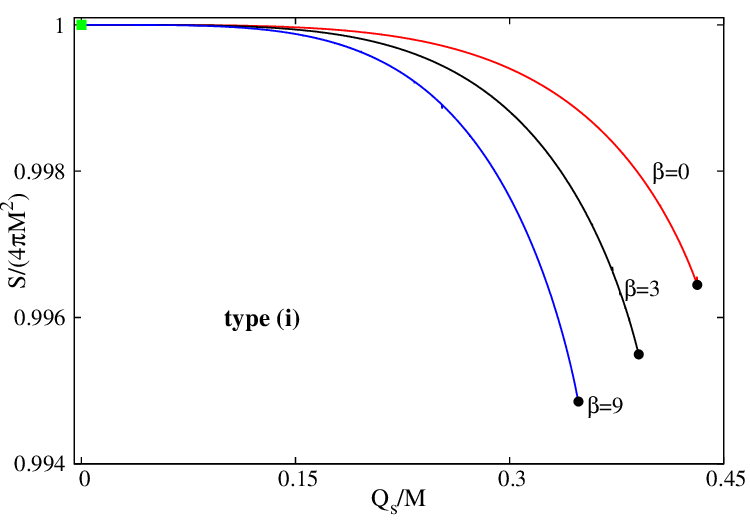} 
      \includegraphics[height=.34\textwidth, angle =0 ]{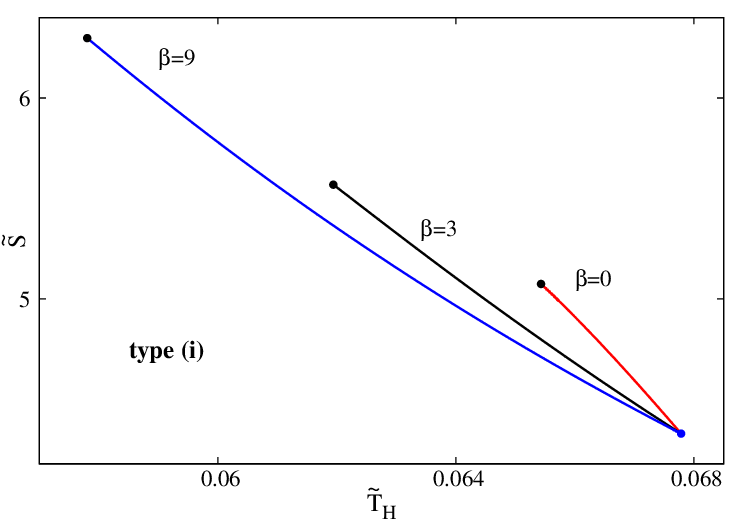} 
 	\end{center}
 	\caption{  
 		{\small 
        {\it Left:} The entropy is shown as a function of 
        scalar charge.
       Both $S$ and $Q_s$ are given in units of mass $M$, with
        $S=4\pi M^2$ for the Schwarzschild BH.
 		       {\it Right:} The entropy is shown as a function of
               Hawking temperature.
 		}
 	}
 	\label{type1-3}
 \end{figure}

A type (i) coupling function with
  $\beta=0$  leads to the scalarized BHs originally found in \cite{Silva:2017uqg,Antoniou:2017acq}.
Solutions with  $\beta>0$ were  reported in Ref.~\cite{Blazquez-Salcedo:2022omw}. 
The fundamental branch of these scalarized BHs is shown  in Fig.~\ref{type1-1}-Fig.~\ref{type1-3}, where we first exhibit their scaled scalar charge $\tilde Q_s$ $vs.$ their scaled mass $\tilde M$ and also $vs.$ their scaled temperature $\tilde T_H$,
for several values of $\beta\geq 0$.
As noted above, for any $\beta$, scalarization sets in at $\tilde M = 0.587$
(marked with a green dot in all plots), 
where the Schwarzschild BH has a zero mode;
for masses $\tilde M> 0.587$, the Schwarzschild BH is linearly stable.
The branch of fundamental scalarized solutions has a relatively short extent (which, however,
increases with $\beta$)
and possesses
a larger mass than the corresponding seed Schwarzschild solution.
This indicates already that it is unstable, as subsequently shown by a linear analysis of the radial mode for $\beta=0$ \cite{Blazquez-Salcedo:2018jnn}
and $\beta >0$  \cite{Silva:2018qhn}. 
Let us remark that the Hawking temperature
decreases along the
branch of fundamental scalarized solutions.
For any $\beta \geq 0$, the branch of scalarized BHs
ends in a critical solution (marked with a black dot in all plots)
 where the radicand $\Delta$ ($cf.$ (\ref{Delta})) vanishes.

Considering next the thermodynamical stability, we exhibit in Fig.~\ref{type1-2} 
the difference between the free energy of  scalarized BHs and those of the 
 vacuum ones as a function of mass and Hawking temperature.
The figure shows that the free energy of these scalarized BHs is always greater than the free energy $F_\text{S}$ of the Schwarzschild BHs.
Clearly, there is no phase transition 
between vacuum and scalarized solutions
for this choice of the coupling functions. 

Moreover, as one can see in 
 Fig.~\ref{type1-3} 
 the entropy of scalarized solutions
 is smaller than the one of a Schwarzschild BH with the same mass 
 (in which case $S=4 \pi M^2$).
 Also, the entropy 
 decreases with temperature, indicating that the scalarized solutions
 possess  a negative specific heat.

\subsubsection{A type (ii) function 
$f(\phi)=\frac{1}{2\beta}[1-\exp(-\frac{\beta\phi^2}{4})]$
}

 \begin{figure}[ht!]
 	\begin{center}
    \includegraphics[height=.34\textwidth, angle =0 ]{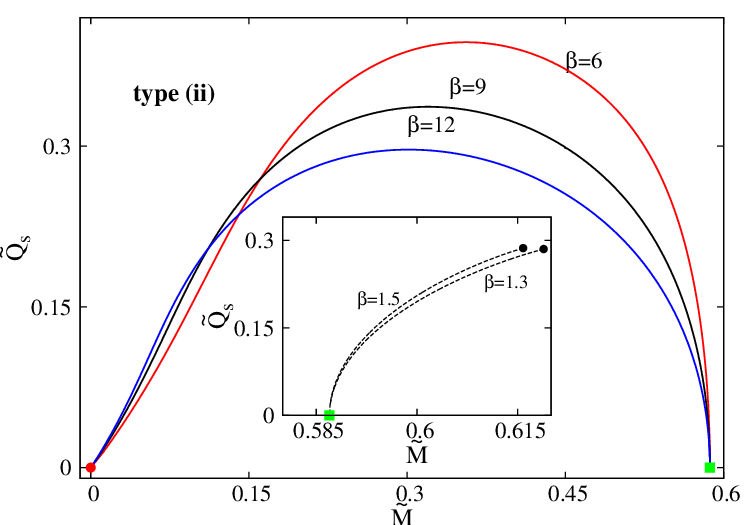}
    \includegraphics[height=.34\textwidth, angle =0 ]{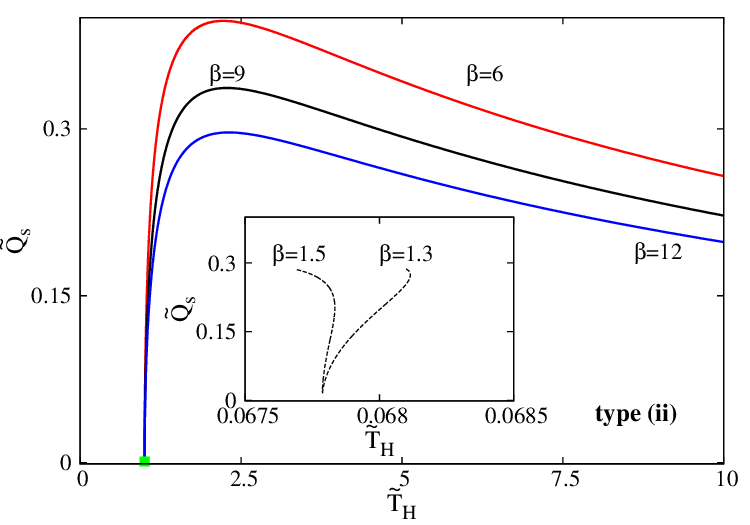} 
 	\end{center}
 	\caption{  
 		{\small 
	Same as Figure \ref{type1-1} for a type (ii)
    coupling function.
 		}
 	}
 	\label{type2-1}
 \end{figure}

 \begin{figure}[ht!]
 	\begin{center}
   \includegraphics[height=.34\textwidth, angle =0 ]{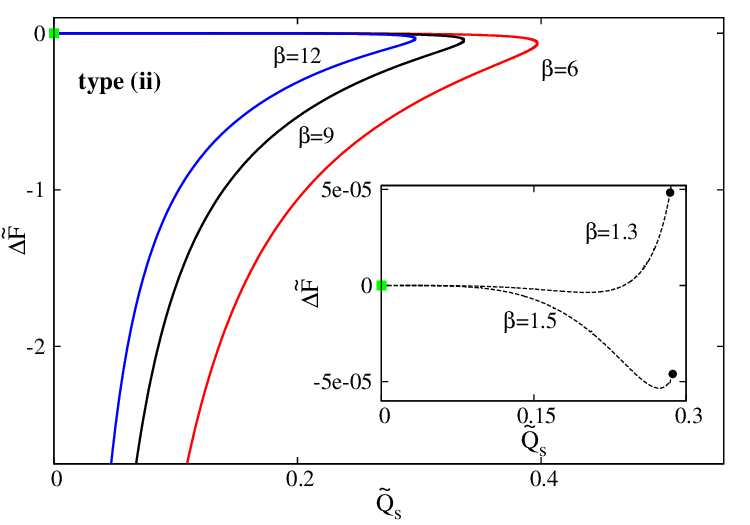}
    \includegraphics[height=.34\textwidth, angle =0 ]{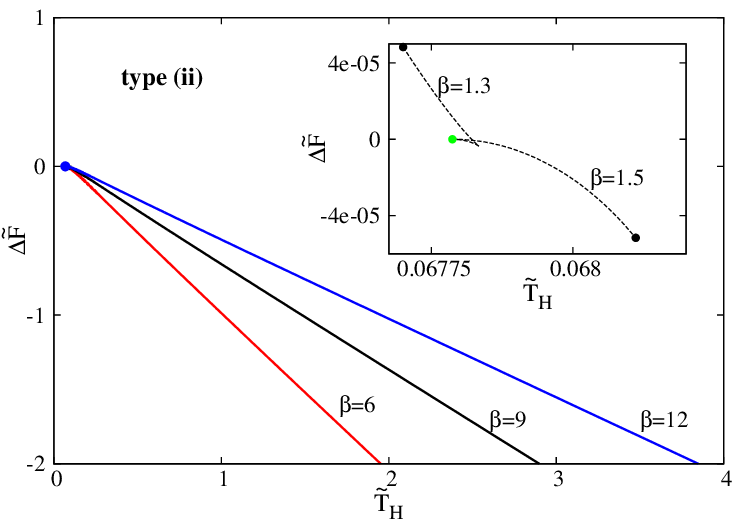} 
 	\end{center}
 	\caption{  
 		{\small 
	Same as Figure \ref{type1-2} for a type (ii)
    coupling function.  
 		}
 	}
 	\label{type2-2}
 \end{figure}

 \begin{figure}[ht!]
 	\begin{center}
    \includegraphics[height=.34\textwidth, angle =0 ]{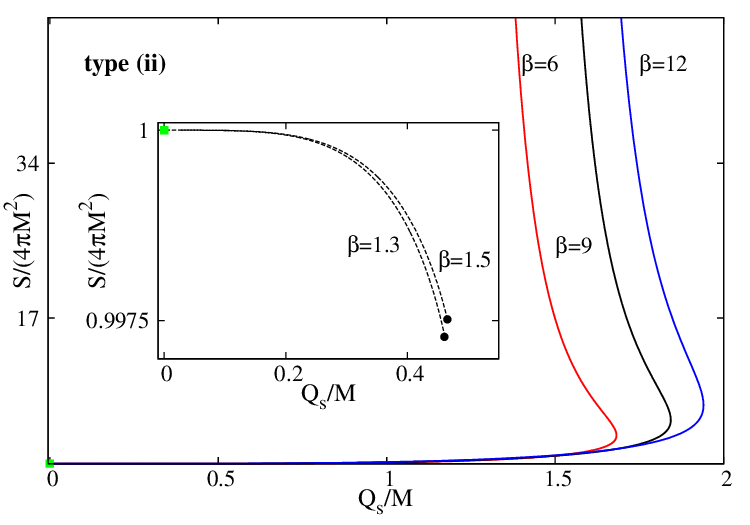} 
      \includegraphics[height=.34\textwidth, angle =0 ]{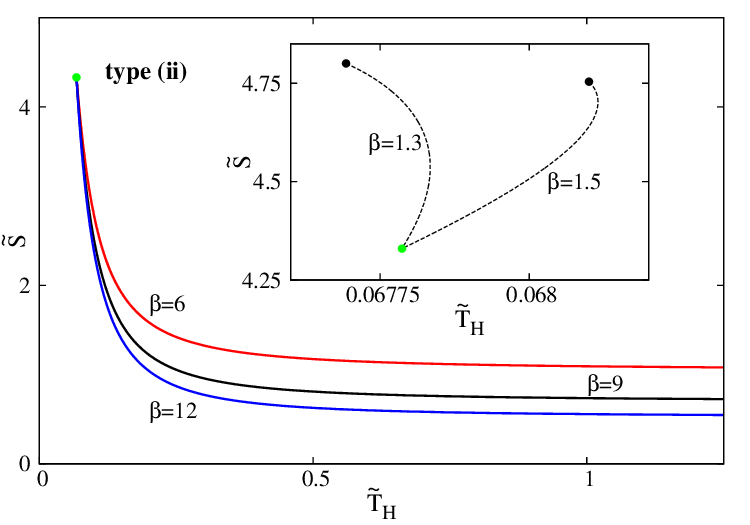} 
 	\end{center}
 	\caption{  
 		{\small 
	Same as Figure \ref{type1-3} for a type (ii)
    coupling function.
 		}
 	}
 	\label{type2-3}
 \end{figure}

We next consider the second coupling function, 
$f(\phi)=\frac{1}{2\beta}[1-\exp(-\frac{\beta\phi^2}{4})]$, 
where $\beta$ is a positive constant \cite{Doneva:2017bvd}. 
The choice of an exponential coupling function appears inspired by the spontaneous scalarization of neutron stars \cite{Damour:1993hw}.
Also, as with a type (i) coupling function,  it
leads to the same set of tachyonic instabilities that give rise to branches of spontaneously scalarized BHs \cite{Doneva:2017bvd}.

  However, a type (ii) function leads to a much richer phase structure.
As before,  no phase transition occurs 
for a range of $\beta$ close to zero.
Yet the picture changes  once the parameter  $\beta$ becomes sufficiently large.
Choosing $e.g.$
$\beta=6$ \cite{Doneva:2017bvd}, the fundamental branch is shown in Fig.~\ref{type2-1}, where we exhibit again the (scaled) scalar charge versus the (scaled) mass
and Hawking temperature.
We notice immediately that, different from the previous case, the fundamental branch bends towards smaller values of the mass
and larger temperatures, which hints towards linear stability.
It extends all the way towards
a limiting singular  solution with vanishing mass  and scalar charge (marked with a red dot in the plots).
Indeed, whereas the Schwarzschild BHs are unstable for masses below $\tilde M=0.587$, the scalarized BHs on the fundamental branch do not possess an unstable radial mode after the bifurcation from the Schwarzschild branch \cite{Blazquez-Salcedo:2018jnn}.
We note, however, that below a critical value of the mass, hyperbolicity is lost.\footnote{ Here, hyperbolicity refers to the well-posedness of the underlying field equations, which is required for a consistent dynamical evolution.} 
This also holds for the low-lying axial and polar modes \cite{Blazquez-Salcedo:2020rhf,Blazquez-Salcedo:2020caw}.
Thus the fundamental branch of scalarized BHs appears to be linearly stable above a certain mass.
The figure also shows the fundamental branches for $\beta=9$ and 12, which show the same characteristics. 
These branches should also be linearly stable according to \cite{Blazquez-Salcedo:2018jnn,Silva:2018qhn}.

 \begin{figure}[ht!]
 	\begin{center}
    \includegraphics[height=.34\textwidth, angle =0 ]{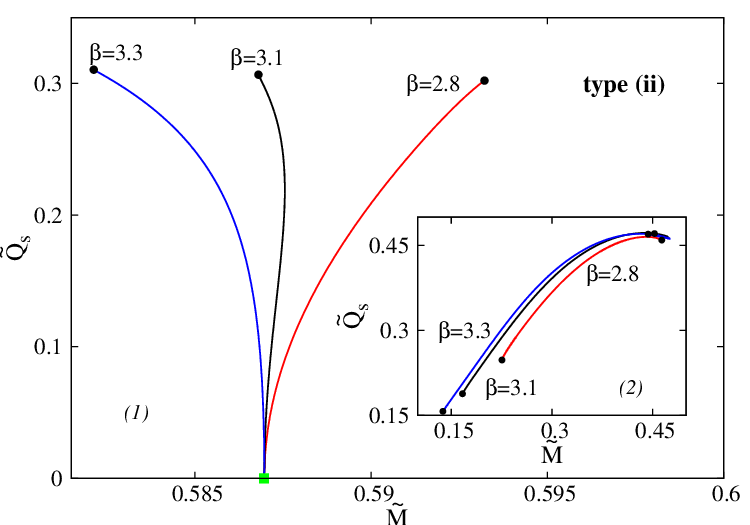}
    \includegraphics[height=.34\textwidth, angle =0 ]{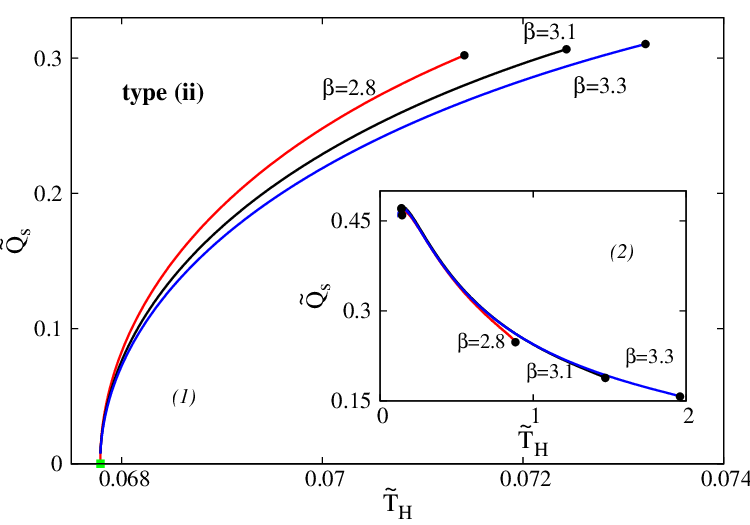} 
 	\end{center}
 	\caption{  
 		{\small 
	Same as Figure \ref{type1-1} for a type (ii)
    coupling function and several values of the coupling constant $\beta$
featuring two disconnected branches of solutions.
 		}
 	}
 	\label{type2t-1}
 \end{figure}

 \begin{figure}[ht!]
 	\begin{center}
   \includegraphics[height=.34\textwidth, angle =0 ]{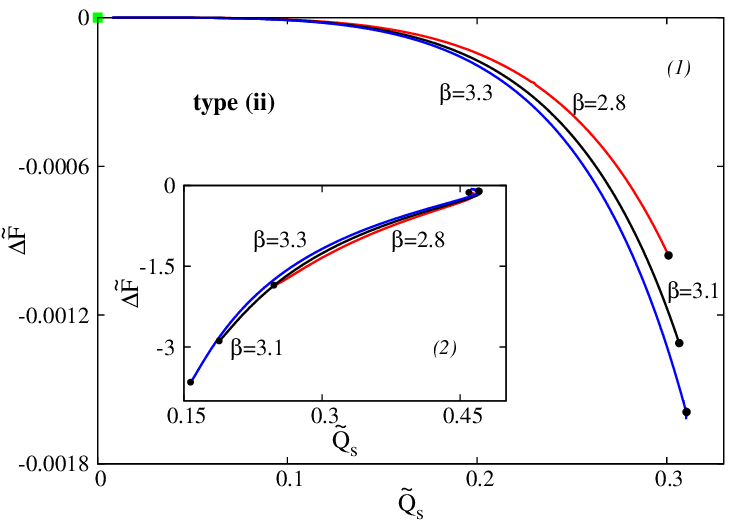}
    \includegraphics[height=.34\textwidth, angle =0 ]{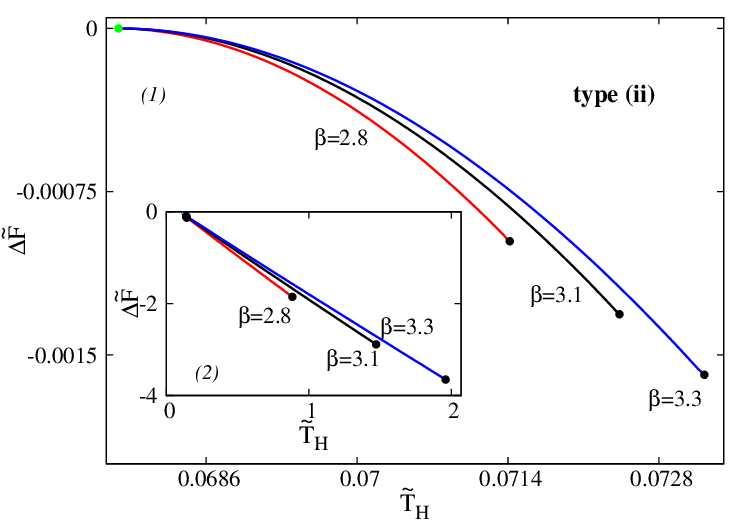} 
 	\end{center}
 	\caption{  
 		{\small 
	Same as Figure \ref{type1-2} for a type (ii)
    coupling function and several values of the coupling constant $\beta$
featuring two disconnected branches of solutions.
 		}
 	}
 	\label{type2t-2}
 \end{figure}

 \begin{figure}[ht!]
 	\begin{center}
    \includegraphics[height=.34\textwidth, angle =0 ]{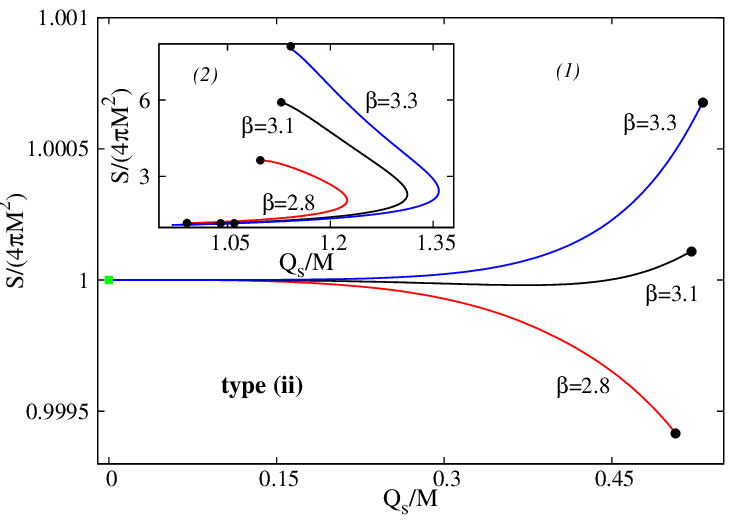} 
      \includegraphics[height=.34\textwidth, angle =0 ]{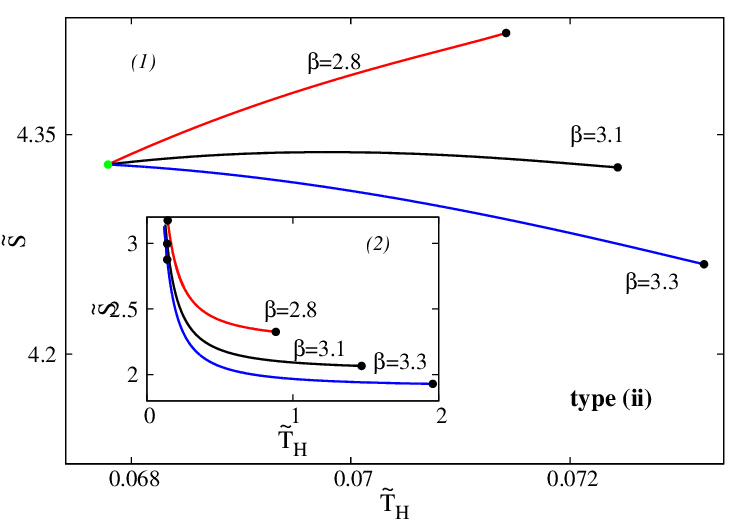} 
 	\end{center}
 	\caption{  
 		{\small 
	Same as Figure \ref{type1-3} for a type (ii)
    coupling function and several values of the coupling constant $\beta$
featuring two disconnected branches of solutions.  
 		}
 	}
 	\label{type2t-3}
 \end{figure}

 \medskip
 
Turning to the free energy of these BHs with $\beta=6,9,12$,
we present in Fig.~\ref{type2-2} the scaled difference $\Delta F = F_\text{EsGB}-F_\text{S}$ of the free energies. 
For these coupling functions, the free energy $F_\text{EsGB}$ of the scalarized BHs is always smaller than the free energy $F_\text{S}$ of the corresponding vacuum GR counterparts.
Thus, in contrast to the polynomial coupling function in the previous subsection, an exponential  coupling functions with sufficiently large $\beta$,
allows for scalarized BHs which are thermodynamically favored over the Schwarzschild ones. 
At the transition point
(with a critical temperature $T_H^{(c)})\simeq 0.0678$), 
the mass and the first derivative of the free energy is continuous.
We conclude that the phase transition from Schwarzschild to scalarized BHs is a smooth \textit{second-order} transition for sufficiently large values of 
$\beta$.  In terms of the expansion~\eqref{le}, this corresponds to 
$b>0$, such that the minimum of $\Delta F$ shifts continuously from $Q_s=0$ to 
$Q_s\neq 0$ as the temperature crosses the critical value.

We remark that the entropy of these scalarized solutions is greater than the entropy of the Schwarzschild solutions with the same mass, see Fig.~\ref{type2-3}.
However, as one can see there, $S$
decreases with the temperature.  Therefore they possess a negative specific heat, a feature inherited from the Schwarzschild seed solution.
 
\medskip

However, the situation changes drastically for small enough values of the coupling constant $\beta$,
with a critical value $\beta \approx 3.2$,
as displayed in Figs.~\ref{type2t-1}-\ref{type2t-3} (insets).  
  In this case, 
the scalarized branches will bend to higher masses when emerging from the bifurcation point (or  towards lower temperatures in a $\tilde Q_s(\tilde T_H)$ plot) -- see the insets in Fig.~\ref{type2t-1}. 
For a small range of values of $\beta \le 3.2$ though in a $\tilde Q_s(\tilde M)$ diagram, a maximal value of $\tilde M$ is approached with a  turning point, and  then the curve bends backward toward smaller $\tilde M$, until it ends in a critical configuration with a vanishing radicand
$\Delta$ as defined in eq.~(\ref{Delta}).

By analogy with other  cases where scalarized solutions reach extremal values of $\tilde  M$ and thus turning points, 
we conjecture that at the turning points a zero mode is encountered, where the radial stability of the scalarized solutions changes (see $e.g.$ ~\cite{Silva:2018qhn,Macedo:2019sem,Blazquez-Salcedo:2022omw}).

However, different from a type (i) coupling function,
the difference in free energies $\Delta F$ takes also negative values.
While  $\Delta F<0$ for  $\beta=1.5$, the situation is more complicated for
$\beta=1.3$, with the existence of two branches together with a cusp,
see the inset in  Fig. \ref{type2-2} (right). 
There
one notices the existence of a {\it first order} phase  
transition, with $\Delta F=0$ for a critical temperature around
$0.06782$ 
corresponding precisely to the maximal value of $\tilde M$, where also linear stability should be gained.
 In the Landau picture, this behavior is associated with $b<0$ and 
$c>0$, leading to the coexistence of multiple local minima of $\Delta F$ and a discontinuous jump of the order parameter. 
 
Another interesting new feature here is that the entropy of some small-$\beta$ solutions increases with temperature, see the inset in Fig. \ref{type2-3}
(right panel) and also the $\beta=2.8$ curve in  Fig. \ref{type2t-3} ( right, main panel). 
 This is a unique feature among the solutions considered in this work, they possessing a positive specific heat, 
 with a well-defined canonical ensemble.

\medskip  

 Another interesting phenomenon occurs for 
$2.4\lesssim  \beta  \lesssim 4.7$ -- namely the scalarized BH solutions consist of two
disconnected branches. 
The existence of a gap between them is
due to the existence criterion of scalarized BHs, eq.~(\ref{Delta}).
The situation here is illustrated in Figs. \ref{type2t-1}- \ref{type2t-3}
for three representative values of $\beta$,
the two branches being marked with $(1)$ and $(2)$.

While the first branch exhibits the pattern
discussed above for small $\beta$,
the branch (2) (shown in the insets) has rather unusual features.
First, it corresponds to {\it non-linear scalarization},
being disconnected from the Schwarzschild solution
(that is, the scalar charge never becomes zero).
Also, 
$\Delta F$ is again negative, the scalarized solutions
being thermodynamically favored
over the Schwarzschild one.
However, there is no critical temperature defined by a crossing of the $F(T_H)$ curves.
Moreover, since the sets (1) and (2) of scalarized solutions
exist for a different range of temperatures,
no phase transition occurs also between them.

Linear radial stability for disconnected branches was studied in \cite{Silva:2018qhn} for the polynomial truncation of the exponential coupling showed no radial instability for set (2) solutions.
By analogy, we conclude that at least parts of these branches may be linearly stable.

\subsection{Non-linear scalarization with $f(\phi)=\frac{1}{4\beta}[1-\exp(-\frac{\beta\phi^4}{16})]$
}

 \begin{figure}[ht!]
 	\begin{center}
    \includegraphics[height=.34\textwidth, angle =0 ]{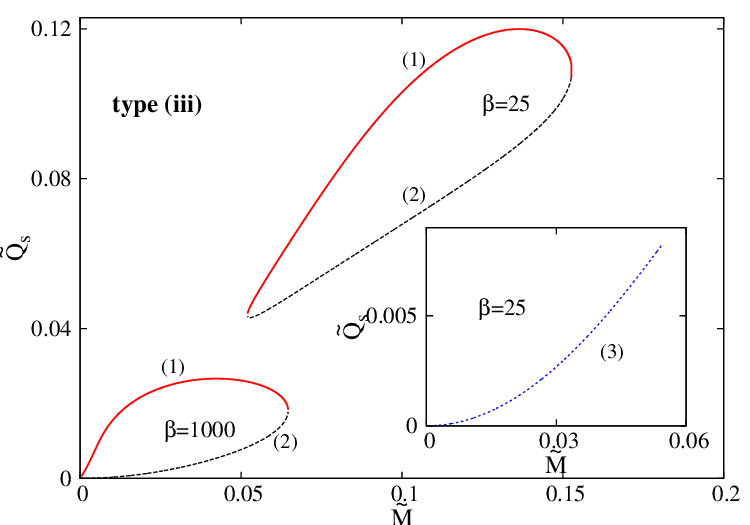}
    \includegraphics[height=.34\textwidth, angle =0 ]{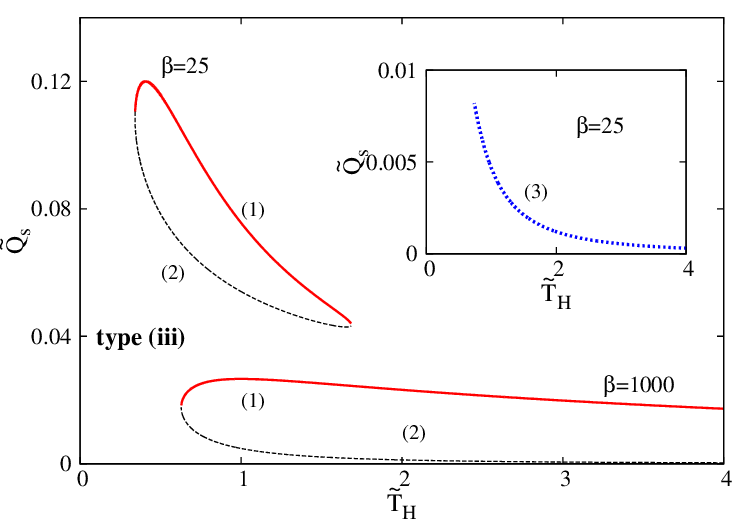} 
 	\end{center}
 	\caption{  
 		{\small 
	Same as Figure \ref{type1-1} for a type (iii)
    coupling function.
 		}
 	}
 	\label{type3-1}
 \end{figure}

 \begin{figure}[ht!]
 	\begin{center}
   \includegraphics[height=.34\textwidth, angle =0 ]{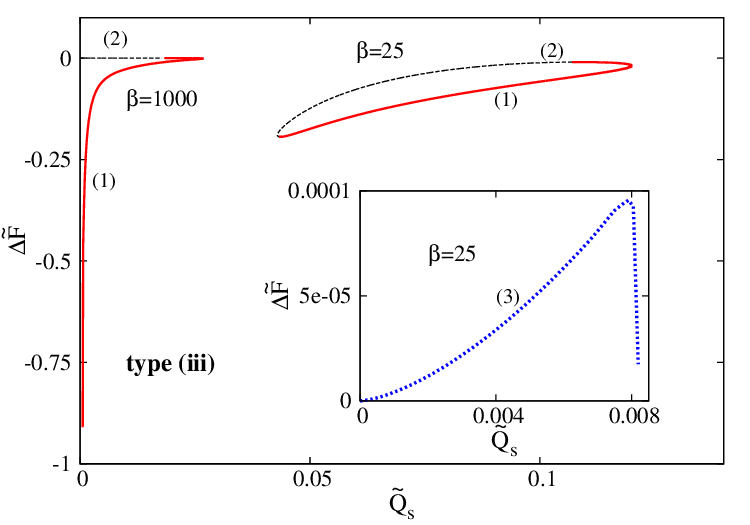}
    \includegraphics[height=.34\textwidth, angle =0 ]{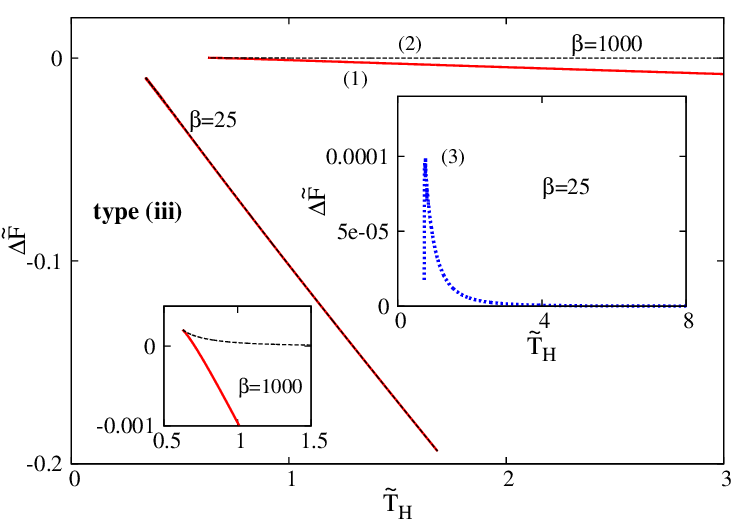} 
 	\end{center}
 	\caption{  
 		{\small 
	Same as Figure \ref{type1-2} for a type (iii)
    coupling function.
 		}
 	}
 	\label{type3-2}
 \end{figure}

 \begin{figure}[ht!]
 	\begin{center}
    \includegraphics[height=.34\textwidth, angle =0 ]{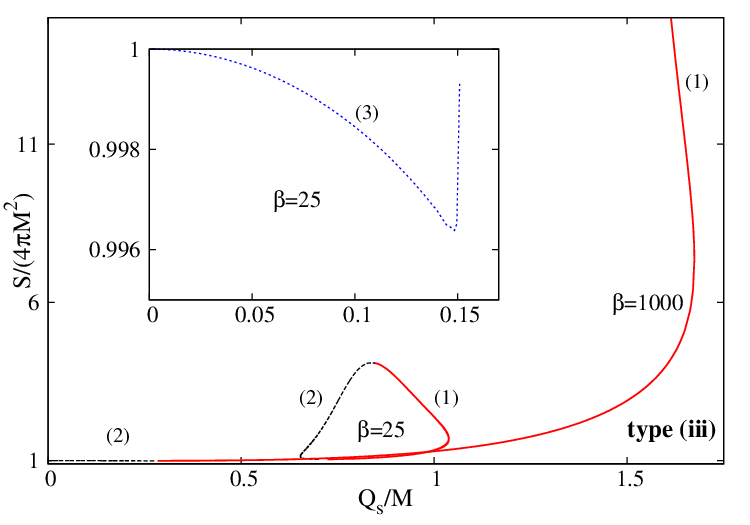} 
      \includegraphics[height=.34\textwidth, angle =0 ]{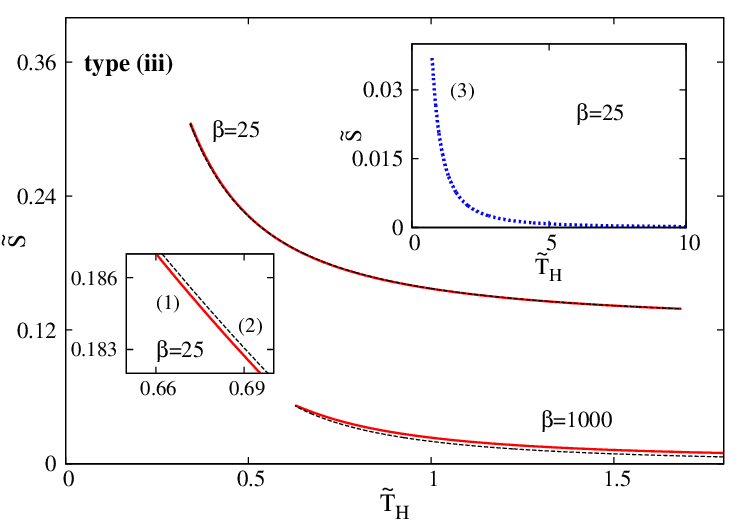} 
 	\end{center}
 	\caption{  
 		{\small 
	Same as Figure \ref{type1-3} for a type (iii)
    coupling function.  
 		}
 	}
 	\label{type3-3}
 \end{figure}

We finally consider the type (iii) coupling function. 
As discussed above, 
  only non-linear scalarization is possible in this case.
These solutions  were first obtained in \cite{Doneva:2021tvn},
their radial stability being investigated in  \cite{Blazquez-Salcedo:2022omw}.

Following the same route as before,
we start by exhibiting in  Fig.~\ref{type3-1} 
the scalar charge of the solutions 
as a function of  mass
and Hawking temperature, for two representative values of $\beta$.
For $\beta=1000$, one notices the existence of two branches of scalarized BHs that emanate from a
singular configuration with zero horizon size and vanishing mass and scalar charge (and also diverging temperature).
Those two branches merge at a maximal value of the mass, forming a closed loop in those diagrams.
Also, the upper branch (solid line marked with (1) in the plots) 
is radially linearly stable above a critical value  where hyperbolicity is lost,
whereas the lower branch (dashed, marked with (2)) is radially unstable \cite{Blazquez-Salcedo:2022omw}.

However, for $\beta=25$, the upper branch (again marked with (1)) 
is no longer connected to the origin, and it is non-hyperbolic.
The lower branch (that is also non-hyperbolic)  merges with the upper branch, forming a closed loop of solutions in the corresponding
($\tilde M, \tilde Q_s$) and ($\tilde T_H, \tilde Q_s$)-diagrams. 

\medskip
 
Turning to the free energy, we exhibit 
 in Fig.~\ref{type3-2}
the (scaled) difference $\Delta F = F_\text{EsGB}-F_\text{S}$ of the free energies as a function of
(scaled) mass and temperature. 
The above mentioned two branch structure 
of scalarized solutions results in a double branch structure of $\Delta F$.
The minimal free energy is found for the first branch configurations.

For $\beta=1000$, the lower scalarized branch (dashed) and the upper scalarized branch (solid) form a ``cusp" at the maximal scalar charge.
The free energy of the lower scalarized branch is always greater than the Schwarzschild free energy.
In contrast, 
$\Delta F< 0$ for the upper scalarized branch (index (1)), 
except close to the cusp, see the inset in Fig. \ref{type3-2} (right).
Therefore there will be a {\it first order} phase transition between the Schwarzschild BH and the non-linear scalarized BHs with $\beta=1000$
at $\delta F=0$. This corresponds to a critical (scaled) temperature $\tilde T_H^{(c)}\simeq 0.70894$.

The picture is somehow different for $\beta=25$, with $\Delta F<0$
for all range of $T_H$.  
In that case
there are no stable scalarized BHs, with the existence of only radially unstable and non-hyperbolic scalarized solutions.
This excludes the existence of
a phase transition between the GR and EsGB BHs.

For both $\beta=25$ and $\beta=1000$
the entropy of a scalarized solution is always 
larger than the one of a Schwarzschild BH with the same mass, see Fig. \ref{type3-3}.
However, these solutions possess again a negative specific heat.

A specific feature found for small enough values of $\beta$
(the choice $\beta=25$ being representative)
is the occurrence of a disconnected set of solutions, marked with $(3)$ in the insets of Figs.
(\ref{type3-1})-(\ref{type3-3}).
This branch  emerges from the origin ($Q_s=M=0$)
and is radially unstable \cite{Blazquez-Salcedo:2022omw}.
Some basic features in this case are similar to those for solutions with
a type (i) coupling.
For example,
$\delta F = F_\text{EsGB}-F_\text{S}$ is always positive,
while they are 
entropically disfavored over the Schwarzschild solution. Moreover, they also possess a negative specific heat, see the insets in Fig. (\ref{type3-3}). 

\section{Conclusions and Outlook}

We have revisited the phenomenon of BH scalarization in EsGB gravity from a thermodynamic perspective and investigated the presence and the order of phase transitions between the Schwarzschild solution of general relativity and scalarized BHs.
Our analysis focused on static, spherically symmetric configurations and on the fundamental scalarized branch, since excited scalarized solutions typically possess radial instabilities.
We have considered three representative classes of scalar-Gauss-Bonnet coupling functions $f(\phi)$, all even in $\phi$, which therefore yield pairs of degenerate scalarized solutions related by the transformation $\phi \rightarrow -\phi$.

Overall, we find a rich phase structure that depends sensitively on the form of the coupling function and on the values of its parameters, including the coexistence of multiple scalarized branches and transitions of different order. In particular, depending on the coupling function and its parameters, three distinct scenarios may arise: the absence of a phase transition, a continuous second–order transition at the scalarization threshold, or a discontinuous first–order transition between Schwarzschild and scalarized BHs.

For the polynomial coupling
$f(\phi)=\tfrac{1}{8}\phi^2+\tfrac{\beta\phi^4}{64}\  (\beta\ge0)$, which represents the simplest extension leading to spontaneous scalarization, the fundamental scalarized branch is radially unstable.
Moreover, the free energy of these scalarized solutions is always larger than that of the Schwarzschild BH for a given temperature.
Consequently, the scalarized configurations are thermodynamically disfavored and no phase transition occurs.

The exponential coupling
$
f(\phi)=\frac{1}{2\beta}\left[1-\exp\left(-\frac{\beta\phi^2}{4}\right)\right]
$
also allows for spontaneous scalarization but leads to a much richer phase structure.
For sufficiently large values of the coupling parameter $\beta$, the fundamental scalarized branch emerging from the bifurcation point is radially stable as the mass decreases from the bifurcation value, until hyperbolicity is lost at sufficiently small masses, whereas the Schwarzschild solution becomes unstable below the bifurcation point.
In this regime the free energy of the scalarized branch is smaller than that of the Schwarzschild solution, yielding a continuous second-order phase transition at the bifurcation point.
For smaller values of $\beta$, however, the scalarized branch develops a turning point where radial stability changes.
This produces a region where both scalarized and Schwarzschild solutions are locally stable.
Since the scalarized configurations possess lower free energy, a discontinuous first-order phase transition occurs.
In addition, a disconnected branch of nonlinearly scalarized solutions may appear, which can be thermodynamically favored over the Schwarzschild solution without being connected to it through a phase transition.

 \begin{figure}[ht!]
 	\begin{center}
    \textit{spontaneous scalarization}\par\medskip
    \includegraphics[height=.34\textwidth, angle =0 ]{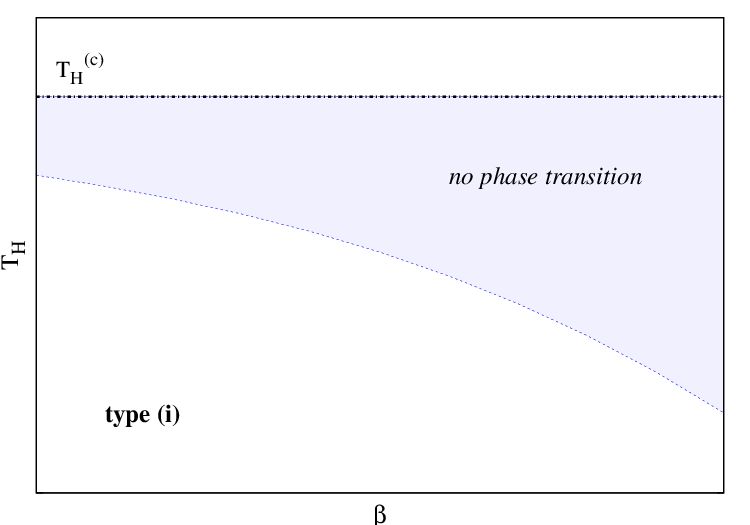} 
      \includegraphics[height=.34\textwidth, angle =0 ]{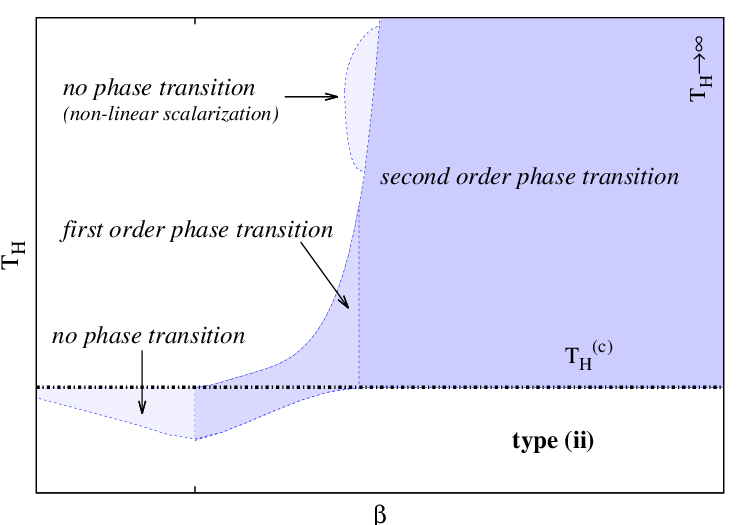} 
 	\end{center}
 	\caption{\small A qualitative diagram summarizing the emerging picture for type (i) and  type (ii)
coupling functions.}
 	\label{type12}
\end{figure}

 \begin{figure}[ht!]
 	\begin{center}
    \textit{non-linear scalarization}\par\medskip
    \includegraphics[height=.34\textwidth, angle =0 ]{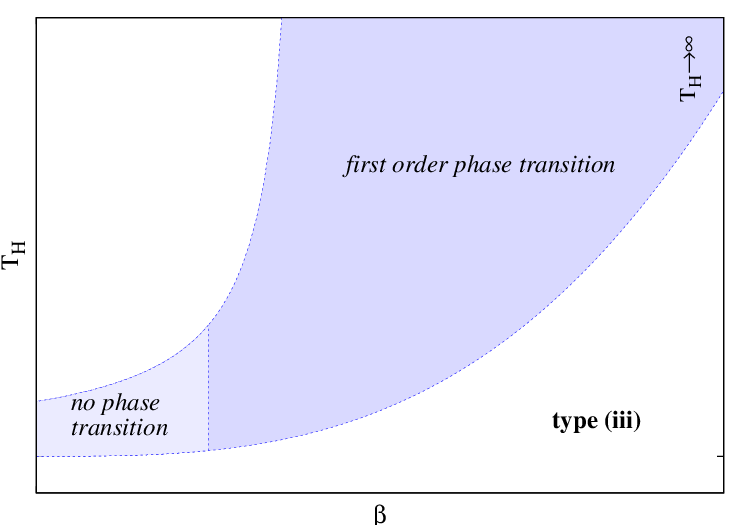}  
 	\end{center}
 	\caption{  
 		{\small Same as Figure \ref{type12} for a type (iii) coupling function. 
 		}
 	}
 	\label{type3}
 \end{figure}

Finally, we considered a coupling function leading exclusively to nonlinear scalarization,
$
f(\phi)=\frac{1}{4\beta}\left[1-\exp\left(-\frac{\beta\phi^4}{16}\right)\right].
$
For large $\beta$, the scalarized solutions form two branches that merge at a maximal mass.
The lower branch is radially unstable and always thermodynamically disfavored relative to the Schwarzschild solution.
The upper scalarized branch becomes radially stable after the cusp where the two branches merge. Along this branch the free energy decreases with decreasing mass and crosses the Schwarzschild free energy at a critical mass. Above this mass the Schwarzschild solution is thermodynamically preferred, whereas below it the scalarized configuration becomes favored, leading to a first-order phase transition.
For sufficiently small $\beta$, on the other hand, only unstable or non-hyperbolic scalarized solutions exist and no phase transition takes place.

Some of these aspects are summarized  
in Figures (\ref{type12}), (\ref{type3}), where we 
display the conjectured domain of existence of the scalarized solutions in 
the $(\beta,T_H)$-plane, in conjunction with the possible existence of a phase transition. 
Those (qualitative) diagrams show  $e.g.$
that, for any $\beta$, the configurations with a polynomial coupling only exist below the critical temperature $T_H^{(c)}$ where the Schwarzschild BH develops a  tachyonic instability.
However, $T_H$ is unbounded for the other two coupling functions with large enough $\beta$.

While our results reveal a diverse thermodynamic phase structure of scalarized BHs, the present work also has several limitations.
First, our analysis is restricted to spherically symmetric solutions.
Rotation is known to play an important role in scalarization, both qualitatively and quantitatively
\cite{Cunha:2019dwb,Collodel:2019kkx,Herdeiro:2020wei,Berti:2020kgk,Doneva:2022yqu}, and extending the present thermodynamic analysis to rotating scalarized BHs would therefore be an important next step.
Second, our discussion of the phase transition relies on thermodynamic arguments and linear stability results available in the literature.
A full understanding of the dynamical evolution between the competing branches would require time-dependent simulations.

In particular, the fate of the scalarized solutions obtained for the polynomial coupling deserves further investigation.
Since these configurations are both radially unstable and thermodynamically disfavored, they are unlikely to represent the final state of the scalarization instability.
Instead, they may correspond to transient configurations appearing along the nonlinear evolution of an unstable Schwarzschild BH before the system relaxes back to the general relativistic branch or to a different scalarized configuration.
Clarifying this issue would require dynamical studies of the nonlinear development of the instability.

Another intriguing aspect concerns the possible analogy between BH scalarization and the Gregory–Laflamme instability of higher-dimensional black strings \cite{Gregory:1993vy}.
In the latter case, the onset of the instability signals the appearance of a new branch of nonuniform solutions that bifurcates from the uniform black string. 
This new branch, however, has less entropy than
the uniform strings and so could not be the putative new
end-state, at least for dimensions lower than thirteen \cite{Sorkin:2004qq}.
Similarly, in EsGB gravity the scalarization instability marks the bifurcation between the Schwarzschild branch and scalarized BH solutions, while for some coupling functions
the scalarized solutions are entropically disfavoured over the vacuum ones.
Exploring this analogy further may help clarify the global structure of the solution space and the role of thermodynamic and dynamical stability in selecting the physically realized configurations.

Several extensions of the present work would be worth pursuing.
One possibility is to include a self-interaction potential for the scalar field \cite{Macedo:2019sem}, which may significantly modify the phase structure. 
It would also be interesting to investigate related scalarization mechanisms in theories with additional curvature couplings \cite{Antoniou:2020nax,Antoniou:2021zoy,Antoniou:2022agj}, where spherically symmetric solutions may be radially stable but unstable in higher multipoles \cite{Kleihaus:2023zzs}.
Finally, analogous phase transitions may arise in theories with different matter content, such as models where the scalar field couples to the Maxwell invariant \cite{Herdeiro:2018wub,Fernandes:2019rez,Astefanesei:2019pfq,Myung:2018vug,Myung:2018jvi,Brihaye:2019kvj,Myung:2019oua,Zou:2019bpt,Blazquez-Salcedo:2020nhs,LuisBlazquez-Salcedo:2020rqp}, or in theories exhibiting spontaneous vectorization \cite{Barton:2021wfj}.

\section*{Acknowledgement}
H. Huang, M.Y. Lai and D.C. Zou gratefully acknowledge support by the National Natural Science Foundation of China (Grant Nos. 12205123, 12565010, 12305064, 1236009). 
E. Radu gratefully acknowledges the support of the Alexander von Humboldt Foundation.
The work of C. Herdeiro and E. Radu is also supported by 
CIDMA (\url{https://ror.org/05pm2mw36}) under the Portuguese Foundation for Science and Technology (FCT, \url{https://ror.org/ 00snfqn58}), Grants UID/04106/2025 (\url{https://doi.org/10.54499/UID/ 04106/2025}) and UID/PRR/04106/2025 (\url{https://doi.org/10.54499/UID/PRR/ 04106/2025}), as well as the projects: Horizon Europe staff exchange (SE) programme HORIZON-MSCA2021-SE-01 Grant No.\ NewFunFiCO-101086251;  2022.04560.PTDC (\url{https://doi.org/10.54499/2022.04560.PTDC}) and 2024.05617.CERN (\url{https://doi.org/10.54499/2024.05617.CERN}).


\end{document}